\documentclass[aps,prl,twocolumn, superscriptaddress, showkeys,longbibliography]{revtex4-2}  

\usepackage{graphicx}
\usepackage{dcolumn}
\usepackage{bm}
\usepackage{hyperref}
\usepackage{graphicx} 
\usepackage{xcolor}
\usepackage{ulem} 

\newcommand{\sqrtsnn}{\ensuremath{\sqrt{s_{\mathrm {NN}}}}}

\newcommand{\gev}{GeV/\ensuremath{c}}
\newcommand{\pp}{\ensuremath{p}+\ensuremath{p}}
\newcommand{\ncoll}{\ensuremath{N_\mathrm{coll}}}
\newcommand{\pT}{\ensuremath{p_\mathrm{T}}}
\newcommand{\pTtrig}{\ensuremath{p_\mathrm{T}^{\mathrm{trig}}}}
\newcommand{\pTjetRaw}{\ensuremath{p_\mathrm{T,jet}^{\mathrm{raw,ch}}}}
\newcommand{\pTjetReco}{\ensuremath{p_\mathrm{T,jet}^{\mathrm{reco,ch}}}}
\newcommand{\pTjet}{\ensuremath{p_\mathrm{T,jet}^{\mathrm{ch}}}}
\newcommand{\pTassoc}{\ensuremath{p_\mathrm{T}^{\mathrm{assoc}}}} 
\newcommand{\dphi}{\ensuremath{\Delta\phi}} 
\newcommand{\epdmip}{\ensuremath{\mathrm{nMIP}_\mathrm{EPD}}}
\newcommand{\Icp}{\ensuremath{I_\mathrm{CP}}}
\newcommand{\IAA}{\ensuremath{I_\mathrm{AA}}}

\newcommand{\antikT}{anti-\ensuremath{k_{\mathrm{T}}}}

\begin{document}

\title{Measurement of jet quenching in O+O collisions \\at $\sqrt{s_\mathrm{NN}}=200$ GeV by the STAR experiment at RHIC}
\affiliation{Academia Sinica, Nankang, 115}
\affiliation{Panchayat College (PC), Bargarh, affiliated with Sambalpur University (SU), Odisha 768028, India}
\affiliation{Abilene Christian University, Abilene, Texas   79699}
\affiliation{Alikhanov Institute for Theoretical and Experimental Physics NRC "Kurchatov Institute", Moscow 117218}
\affiliation{Argonne National Laboratory, Argonne, Illinois 60439}
\affiliation{American University in Cairo, New Cairo 11835, Egypt}
\affiliation{Ball State University, Muncie, Indiana, 47306}
\affiliation{Brookhaven National Laboratory, Upton, New York 11973}
\affiliation{University of Calabria \& INFN-Cosenza, Rende 87036, Italy}
\affiliation{University of California, Berkeley, California 94720}
\affiliation{University of California, Davis, California 95616}
\affiliation{University of California, Los Angeles, California 90095}
\affiliation{University of California, Riverside, California 92521}
\affiliation{Central China Normal University, Wuhan, Hubei 430079 }
\affiliation{University of Illinois at Chicago, Chicago, Illinois 60607}
\affiliation{Chongqing University, Chongqing, 401331}
\affiliation{Creighton University, Omaha, Nebraska 68178}
\affiliation{Czech Technical University in Prague, FNSPE, Prague 115 19, Czech Republic}
\affiliation{National Institute of Technology Durgapur, Durgapur - 713209, India}
\affiliation{ELTE E\"otv\"os Lor\'and University, Budapest, Hungary H-1117}
\affiliation{Frankfurt Institute for Advanced Studies FIAS, Frankfurt 60438, Germany}
\affiliation{Fudan University, Shanghai, 200433 }
\affiliation{Guangxi Normal University, Guilin, 541004}
\affiliation{University of Heidelberg, Heidelberg 69120, Germany }
\affiliation{University of Houston, Houston, Texas 77204}
\affiliation{Huzhou University, Huzhou, Zhejiang  313000}
\affiliation{Indian Institute of Science Education and Research (IISER), Berhampur 760010 , India}
\affiliation{Indian Institute of Science Education and Research (IISER) Tirupati, Tirupati 517507, India}
\affiliation{Indian Institute Technology, Patna, Bihar 801106, India}
\affiliation{Indiana University, Bloomington, Indiana 47408}
\affiliation{Institute of Modern Physics, Chinese Academy of Sciences, Lanzhou, Gansu 730000 }
\affiliation{University of Jammu, Jammu 180001, India}
\affiliation{Joint Institute for Nuclear Research, Dubna 141 980}
\affiliation{Kent State University, Kent, Ohio 44242}
\affiliation{University of Kentucky, Lexington, Kentucky 40506-0055}
\affiliation{Lanzhou University, Lanzhou, 730000}
\affiliation{Lawrence Berkeley National Laboratory, Berkeley, California 94720}
\affiliation{Lehigh University, Bethlehem, Pennsylvania 18015}
\affiliation{Lovely Professional University, Jalandhar - Delhi G.T. Road, Pagwara, Panjab, 144411, India}
\affiliation{Max-Planck-Institut f\"ur Physik, Munich 80805, Germany}
\affiliation{Michigan State University, East Lansing, Michigan 48824}
\affiliation{National Research Nuclear University MEPhI, Moscow 115409}
\affiliation{National Institute of Science Education and Research, HBNI, Jatni 752050, India}
\affiliation{National Cheng Kung University, Tainan 70101 }
\affiliation{The Ohio State University, Columbus, Ohio 43210}
\affiliation{Panjab University, Chandigarh 160014, India}
\affiliation{NRC "Kurchatov Institute", Institute of High Energy Physics, Protvino 142281}
\affiliation{Purdue University, West Lafayette, Indiana 47907}
\affiliation{Rice University, Houston, Texas 77251}
\affiliation{Rutgers University, Piscataway, New Jersey 08854}
\affiliation{University of Science and Technology of China, Hefei, Anhui 230026}
\affiliation{South China Normal University, Guangzhou, Guangdong 510631}
\affiliation{Sejong University, Seoul, 05006, Korea, Republic Of}
\affiliation{Shandong University, Qingdao, Shandong 266237}
\affiliation{Shanghai Institute of Applied Physics, Chinese Academy of Sciences, Shanghai 201800}
\affiliation{Southern Connecticut State University, New Haven, Connecticut 06515}
\affiliation{State University of New York, Stony Brook, New York 11794}
\affiliation{Instituto de Alta Investigaci\'on, Universidad de Tarapac\'a, Arica 1000000, Chile}
\affiliation{Temple University, Philadelphia, Pennsylvania 19122}
\affiliation{Texas A\&M University, College Station, Texas 77843}
\affiliation{Texas Southern University, Houston, Texas, 77004}
\affiliation{University of Texas, Austin, Texas 78712}
\affiliation{Tsinghua University, Beijing 100084}
\affiliation{University of Tsukuba, Tsukuba, Ibaraki 305-8571, Japan}
\affiliation{University of Chinese Academy of Sciences, Beijing, 101408}
\affiliation{Valparaiso University, Valparaiso, Indiana 46383}
\affiliation{Variable Energy Cyclotron Centre, Kolkata 700064, India}
\affiliation{Warsaw University of Technology, Warsaw 00-661, Poland}
\affiliation{Wayne State University, Detroit, Michigan 48201}
\affiliation{Wuhan University of Science and Technology, Wuhan, Hubei 430065}
\affiliation{Yale University, New Haven, Connecticut 06520}

\author{B.~E.~Aboona}\affiliation{Texas A\&M University, College Station, Texas 77843}
\author{J.~Adam}\affiliation{Czech Technical University in Prague, FNSPE, Prague 115 19, Czech Republic}
\author{G.~Agakishiev}\affiliation{Joint Institute for Nuclear Research, Dubna 141 980}
\author{I.~Aggarwal}\affiliation{Panjab University, Chandigarh 160014, India}
\author{M.~M.~Aggarwal}\affiliation{Panjab University, Chandigarh 160014, India}
\author{Z.~Ahammed}\affiliation{Variable Energy Cyclotron Centre, Kolkata 700064, India}
\author{A.~Aitbayev}\affiliation{Joint Institute for Nuclear Research, Dubna 141 980}
\author{I.~Alekseev}\affiliation{Alikhanov Institute for Theoretical and Experimental Physics NRC "Kurchatov Institute", Moscow 117218}\affiliation{National Research Nuclear University MEPhI, Moscow 115409}
\author{E.~Alpatov}\affiliation{National Research Nuclear University MEPhI, Moscow 115409}
\author{A.~K.~Alshammri}\affiliation{Kent State University, Kent, Ohio 44242}
% ...

\collaboration{STAR Collaboration}\noaffiliation

\begin{abstract}
The STAR experiment at the Relativistic Heavy Ion Collider presents measurements of correlations between charged hadron triggers of high transverse momenta ($7 < p_{\rm T} < 30$ GeV/$c$) with recoiling charged hadrons ($3 < p_{\rm T} < 7$ GeV/$c$) or charged--particle jets ($p_{\rm T, jet} > 8$ GeV/$c$) in event--activity selected O+O collisions at $\sqrtsnn=200$ GeV. Yields of associated hadrons and jets, normalized by the number of trigger hadrons, are suppressed by approximately 20\% in high event activity relative to low event activity collisions, with an absence of suppression excluded with high significance. This suppression corresponds to a shift in \pT\ of $0.70\pm0.15~(\rm stat.)~\pm0.10~(\rm syst.)$ GeV/$c$ for large--radius charged--particle jets ($R=0.5$), quantifying their energy redistribution due to final--state interactions. These measurements provide strong evidence for jet quenching in O+O collisions at $\sqrt{s_\mathrm{NN}}=200$ GeV, offering new insight into quark--gluon plasma formation in small collision systems. 
\end{abstract}

\maketitle

Quark--gluon plasma (QGP) is a state of matter composed of unbound quarks and gluons (collectively called ``partons'')~\cite{Busza:2018rrf,Elfner:2022iae}, with interactions governed by Quantum Chromodynamics (QCD). QGP filled the universe a few microseconds after the Big Bang, and is recreated and studied in the laboratory using high-energy collisions of heavy nuclei at the Relativistic Heavy Ion Collider (RHIC)~\cite{STAR:2005gfr,PHENIX:2004vcz,PHOBOS:2004zne,BRAHMS:2004adc} and the Large Hadron Collider (LHC)~\cite{ALICE:2022wpn,CMS:2024krd}. 

Analysis of heavy--ion data reveals long--range correlations among particles with low transverse momenta ($p_{\rm T}$ $\lesssim$ $2-3$~\gev), which are well described by hydrodynamic models of a nearly perfect QGP liquid~\cite{Gale:2013da,Heinz:2013th,Heinz:2024jwu}. Additional key evidence for QGP formation is based on measurements of jets~\cite{Harris:2023tti}, which are collimated sprays of hadrons originating from hard (large momentum transfer) scatterings of partons~\cite{Sterman:1977wj,Dasgupta:2014yra,Sapeta:2015gee}. Jets traverse the QGP and interact with it, generating observable modification in jet production rates, correlations, and structure (``jet quenching'')~\cite{Cunqueiro:2021wls,Apolinario:2022vzg,Wang:2025lct,2025-1834}. 

Exploration of the collision system--size dependence of QGP formation can elucidate the fundamental multi--body QCD processes of isotropization, thermalization, and the onset of collective hydrodynamic behavior~\cite{Grosse-Oetringhaus:2024bwr,He:2024xtk}. Long--range correlations observed in the collision of Au+Au and Pb+Pb nuclei~\cite{STAR:2000ekf,ALICE:2010suc}, which have been traditionally ascribed to QGP formation, are also observed in much smaller collision systems, such as \pp, $p$+A, and $e^{+}$+$e^{-}$~\cite{CMS:2010ifv,CMS:2012qk,ATLAS:2017rtr,ALICE:2019zfl,ALICE:2023ulm,PHENIX:2018lia,STAR:2022pfn,STAR:2015kak,ATLAS:2021jhn,Chen:2023njr}. However, no evidence of jet quenching signatures in these small systems has been observed to date~\cite{ALICE:2012mj,CMS:2016xef,ATLAS:2014cpa,ATLAS:2016xpn,ATLAS:2014cpa,ATLAS:2023zfx,PHENIX:2023dxl,ALICE:2017svf,ATLAS:2022iyq,CMS:2025jbv,STAR:2014qsy,STAR:2024nwm}.

Measurements of inclusive hadron and jet production in small systems require the calculation of a yield scaling factor (\ncoll, number of binary collisions) using Glauber modeling~\cite{Miller:2007ri, Li:2020vrg} to account for nuclear geometric effects, based on the correlation of experimentally observed event activity (EA, total multiplicity or energy) and impact parameter~\cite{Miller:2007ri,dEnterria:2020dwq}. In small systems this correlation has been observed to be weak, however, resulting in large systematic uncertainties in the determination of \ncoll\ that limit the sensitivity of such inclusive observables to jet quenching signals~\cite{Alvioli:2017wou,Behera:2023oxe,Perepelitsa:2024eik,Park:2025mbt,Loizides:2025ule} . 

Improved systematic precision in the search for jet quenching signals can be achieved using correlation observables~\cite{STAR:2006vcp,PHENIX:2024twd,ALICE:2011gpa,ALICE:2015mdb,STAR:2017hhs,STAR:2023pal,STAR:2023ksv,ATLAS:2022iyq,He:2024rcv}. In semi--inclusive correlation measurements, where the number of hadrons or jets recoiling from a high--\pT\ ``trigger'' particle (hadron, photon, jet) is counted, the per--trigger yield corresponds to a ratio of two hard production cross sections~\cite{ALICE:2015mdb,STAR:2017hhs}. This ratio has no dependence on Glauber modeling, since the \ncoll\ factors in its numerator and denominator cancel identically~\cite{ALICE:2017svf}. Furthermore, effects arising from modifications to parton distribution functions (PDFs) in nuclei relative to those in unbound nucleons also partially cancel for semi--inclusive observables~\cite{Gebhard:2024flv}. Correlation measurements in small systems have to date likewise not reported a significant jet quenching signal, but their reduced experimental uncertainties enable the determination of an upper limit of 0.4 GeV/$c$~\cite{ALICE:2017svf} or 1.26--1.4\%~\cite{ATLAS:2022iyq,CMS:2025jbv} at 90\% confidence to jet energy loss in high-EA $p$+Pb collisions at the LHC. Further progress in exploring the system-size limits of QGP formation requires measurements of light ion collisions, where jet quenching effects may be larger than achievable experimental uncertainties~\cite{Sievert:2019zjr,Katz:2019qwv,Huss:2020dwe,Huss:2020whe,Zakharov:2021uza,Ke:2022gkq,CMS:2025bta,CMS:2026qef, Wang:2021ghq, Liu:2021izt}. 

In this Letter, the STAR Collaboration reports the first jet quenching measurements in O+O collisions at the center-of-mass energy per nucleon-nucleon pair (\sqrtsnn) of 200 GeV, where long--range correlations have been observed~\cite{STAR:2025ivi}. Charged di--hadron correlations (h-h) and the semi--inclusive distribution of charged--particle jets recoiling from a trigger hadron (h-jet) are reported. Jet quenching effects are assessed by comparing the h-h and h-jet distributions in event populations with small and large EA. 

STAR \cite{STAR:2002eio} is a general--purpose detector at RHIC. Its central barrel has a solenoidal magnetic field of 0.5 Tesla, containing a Time Projection Chamber (TPC) of radius 200 cm that measures charged--particle momenta within pseudo--rapidity acceptance $|\eta|<1.5$ over the full azimuth ($\phi$)~\cite{Anderson:2003ur,YANG:2019rfi}. The dataset for this analysis, recorded in 2021 with a minimum--bias (MB) trigger, comprises approximately 500 million O+O collisions at $\sqrtsnn=200$ GeV, corresponding to an integrated luminosity of 0.14 nb$^{-1}$. The MB trigger requires signals in both the Time--Of--Flight detector, covering $|\eta|<0.9$~\cite{Llope:2003ti}, and forward trigger detectors~\cite{Adler:2000bd,Llope:2014nva,Adams:2019fpo}.

Event vertices, which are reconstructed using charged--particle tracks measured in the TPC, are required to lie within 30 cm in the $z$ direction ($v_{z}$) and within 2 cm radially relative to the center of the TPC. TPC tracks must contain at least 15 TPC hits. The ratio of the number of TPC hits to the maximum possible number along the track trajectory must be greater than 0.52, to suppress track splitting. The three--dimensional distance of closest approach (DCA) of a track to the event vertex must be less than 1 cm. In addition, a selection on the signed transverse DCA is applied, where the sign is defined by the relative orientation of the helical track trajectory and the vertex position, to suppress low--\pT\ particles being erroneously reconstructed with high \pT. Specifically, the signed DCA must be less than 0.5 or greater than -0.5 cm for positively and negatively charged tracks. Accepted tracks have $|\eta|<1.5$ and $\pT>0.2$ \gev.

\begin{figure}[btph]
\begin{center}
\includegraphics[width=0.9\columnwidth, trim=10 15 40 20, clip]{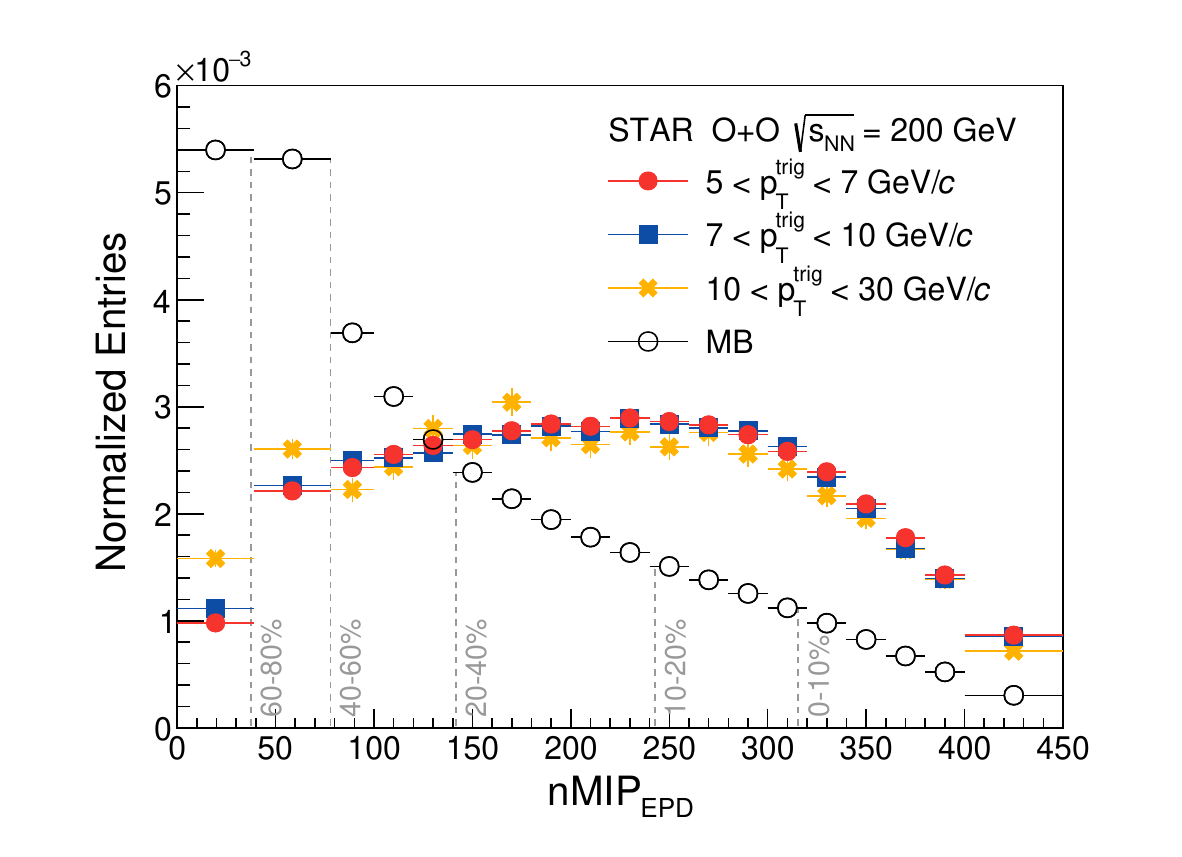}
\caption{Self--normalized distributions of the raw number of MIPs measured in the EPD for MB O+O events (open symbols) and those containing a high--\pT\ trigger particle (filled symbols). Vertical bars represent statistical errors and are mostly smaller than the marker size. Vertical dashed lines indicate selected percentile intervals in EA.}
\label{Fig:EA}
\end{center}
\end{figure}

Event activity is quantified by the number Minimum Ionizing Particles (MIPs) detected in the forward Event Plane Detector (EPD), which covers $2.14<|\eta|<5.09$ over the full azimuth~\cite{Adams:2019fpo}. This choice of EA acceptance differs from the TPC acceptance where hard processes are measured, thereby suppressing self--correlations. Figure~\ref{Fig:EA} shows distributions of the raw number of MIPs measured in the EPD (\epdmip) for the O+O MB population after correcting for the MB trigger efficiency. Vertical dashed lines indicate limits in percentile intervals of the MB distribution used to classify EA, e.g., 0--10\% corresponds to the highest--EA events that make up 10\% of the total hadronic cross section. 

Figure~\ref{Fig:EA} also shows \epdmip\ distributions for event populations containing a high--\pT\ charged particle in the central barrel, in selected \pTtrig\ intervals. These distributions are seen to be biased towards high EA relative to the MB distribution, with the same bias observed for all values of $\pTtrig>5$ \gev\ in the 0--60\% EA percentile interval. See \epdmip\ distribution ratios in the End Matter. Jet production can affect such distributions at low EA since jets recoiling from a high--\pT\ trigger can occur in the EPD acceptance. However, the invariance of the bias with variation in $\pTtrig$ within the 0-60\% interval is consistent with a correlation between EA and collision geometry, with negligible contribution from forward jet production or other long--range correlations. This analysis therefore utilizes the 40-60\% EA interval as the reference distribution, and jet quenching is quantified by \Icp, the ratio of hadron or jet yield measured in high--EA populations to that in the reference EA interval.

\begin{figure}[btph]
\begin{center}
\includegraphics[width=0.8\columnwidth, trim=0 30 25 45, clip]{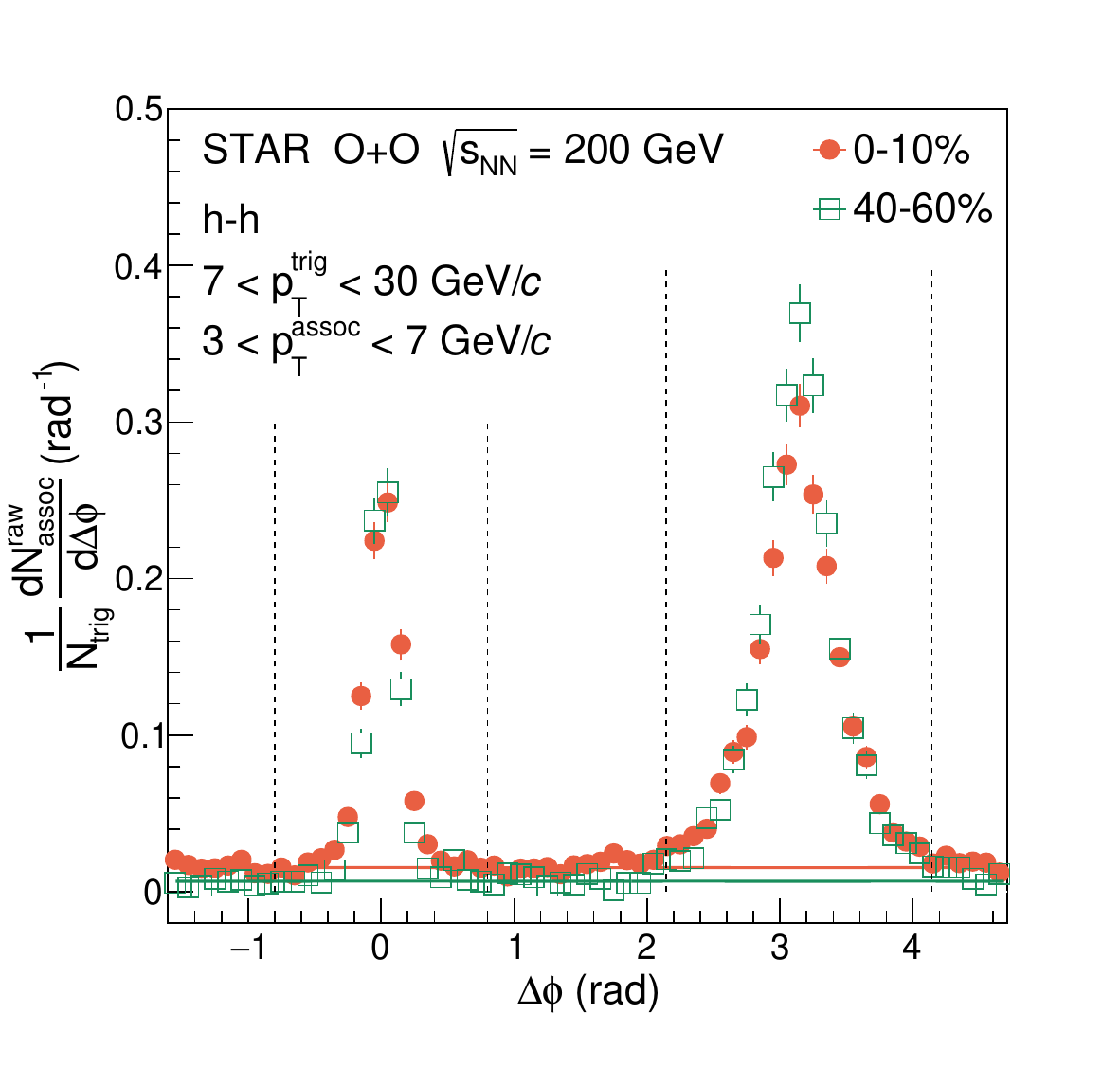}
\includegraphics[width=0.9\columnwidth, trim=11 60 7 40, clip]{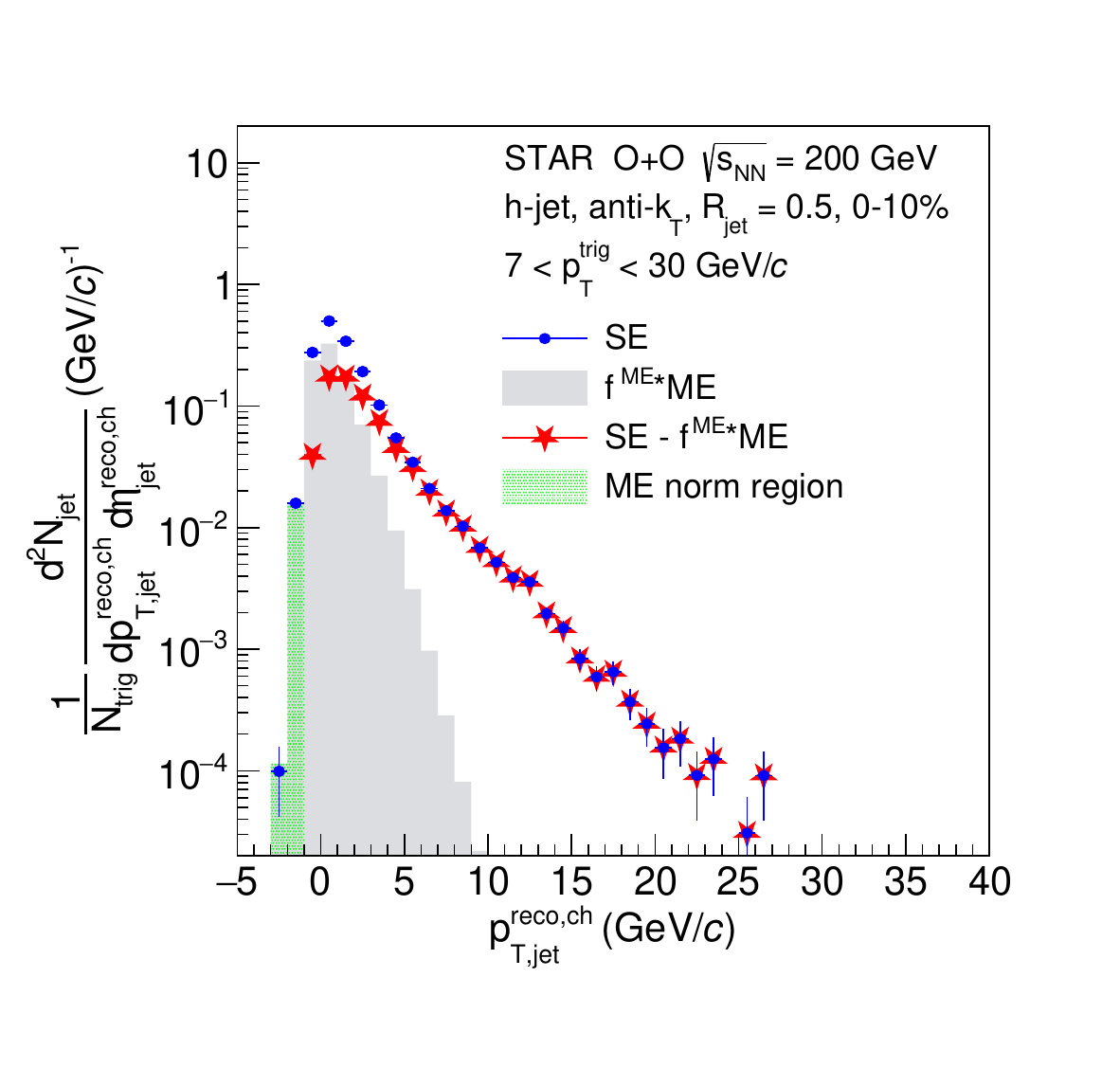}
\caption{Top: raw di--hadron correlations as a function of azimuthal separation $\Delta\phi=\phi^{\rm trig}-\phi^{\rm assoc}$. Solid curves (orange 0-10\%, green 40-60\%) show background fit within $0.8 < |\Delta\phi| < 1.6$, and the vertical lines show signal counting ranges. Bottom: distribution of recoil charged--particle jets as a function of \pTjetReco. The gray filled histogram represents a similar distribution from mixed events as an estimate of combinatorial background, and the green shaded area denotes the range where the ME distribution is normalized. Red stars are the difference between the SE and ME distributions.}
\label{Fig:dihadron}
\end{center}
\end{figure}

Figure \ref{Fig:dihadron}, upper panel, shows raw di--hadron correlations for the 0-10\% and 40-60\% populations, as a function of azimuthal separation $\Delta\phi=\phi^{\rm trig}-\phi^{\rm assoc}$ between a charged hadron trigger with $7<$\pTtrig$<30$ \gev\ and associated charged particles $3<$\pTassoc$<7$ \gev, both within $|\eta|<1.5$. If an event contains multiple tracks satisfying the trigger condition, one track is selected at random and the others are discarded, to ensure that the trigger distribution corresponds to that of inclusive production~\cite{ALICE:2015mdb,STAR:2017hhs}. Two distinct peaks are observed, for small azimuthal separation ($\Delta\phi\sim0$) and back--to--back ($\Delta\phi\sim\pi$), characteristic of fragmentation products of a di--jet pair. 

The $\Delta\phi$ distribution may include yield from associated particles that do not arise from the same hard scattering as the trigger particle. This uncorrelated background yield is estimated by fitting the correlation within $0.8 < |\Delta\phi| < 1.6$ using the functional form  $A(1+2\sum_{n=2}^{4}v_{n}^{2}\cos(n\Delta\phi))$, where $A$ is a normalization factor, and the coefficients $v_{n}$ account for long--range correlations, with values $v_{2}=0.05$, $v_{3} = 0.05$, $v_{4}=v_{2}^{2}$ taken as conservative estimates for all EA classes~\cite{STAR:2025ivi}. The near--side and recoil--side correlated associated yields ($Y_{\rm assoc}$) are determined by integrating the correlation function within $|\Delta\phi|<0.8$ and $|\Delta\phi-\pi|<1$, respectively, and subtracting the uncorrelated background.

The semi--inclusive yield of charged jets recoiling from a high--\pT\ charged hadron trigger is also reported. Charged hadron triggers are selected in the same way as the di--hadron correlation triggers. Charged--particle jet reconstruction utilizes the \antikT\ algorithm \cite{Cacciari:2008gp} implemented in the FastJet package \cite{Cacciari:2005hq,Cacciari:2011ma} with jet radii $R=0.2$ and 0.5. The recoil jet acceptance is $|\eta_{\rm jet}|<1.5-R$ and $\dphi=|\phi^{\rm trig}-\phi^{\rm jet}-\pi|<\pi/4$.

Corrections to the recoil jet \pT\ spectrum for uncorrelated background yield and \pT-smearing are carried out in several steps~\cite{STAR:2017hhs,STAR:2023pal,STAR:2023ksv}. First, the event--wise uncorrelated background density, $\rho$, is estimated via $\rho=\text{median}({p_{\mathrm{ T,jet,} i}^{\rm raw, ch}}/{A_{i}^{\rm jet}})\cdot C$, excluding the two highest--\pT\ jet candidates~\cite{ALICE:2015mdb,STAR:2017hhs}. Here, $p_{\mathrm{ T,jet,} i}^{\rm raw, ch}$ is the raw transverse momentum and $A_{i}^{\rm jet}$ is the area of the $i$-th jet in the event found with the $k_{\rm T}$ algorithm \cite{Catani:1993hr,Ellis:1993tq}. The factor $C$, which accounts for event occupancy, is the ratio of the area covered by all $k_{\rm T}$ jets to the full detector acceptance~\cite{CMS:2012rmf}. The value of \pTjetRaw\ determined by the \antikT\ algorithm is then adjusted according to $\pTjetReco=\pTjetRaw-\rho \cdot A_{\rm jet}$.

Event mixing is then used to estimate the yield of combinatorial jets in the accepted population~\cite{STAR:2017hhs,STAR:2023pal,STAR:2023ksv}. A mixed--event (ME) dataset is constructed from the real data (SE, or ``same event'') utilizing tracks with $\pT<5$ \gev. Mixing is performed separately in distinct subsets of the SE population with similar vertex $z$ (4 bins), event plane orientation (2 bins), and multiplicity distributions (2 bins). Each ME contains at most one track from a given SE. The trigger direction $\phi$ in ME events is chosen randomly. Jet finding is performed on each ME event using the same procedures as the analysis of real data. The resulting ``jet'' population is purely combinatorial, since event mixing suppresses all multi--hadron correlations.

Figure \ref{Fig:dihadron}, lower panel, shows the trigger--normalized \pTjetReco\ distributions of recoil jets in the O+O 0-10\% SE and ME event populations. The negative \pTjetReco\ region in the SE distribution is expected to arise predominantly from combinatorial jets that are uncorrelated with the trigger, indicated by the similarity in the shape of the SE and ME distributions in that region, while the SE yield should be larger at large positive values of \pTjetReco\ due to jet candidates correlated with the trigger~\cite{ALICE:2015mdb,STAR:2017hhs,STAR:2023pal,STAR:2023ksv}. The ME distribution is therefore normalized to SE in the background--dominated \pTjetReco\ region with a factor $f^{\rm ME}$, providing the \pTjetReco\ distribution of combinatorial jet background yield~\cite{ALICE:2015mdb,STAR:2017hhs,STAR:2023pal,STAR:2023ksv}, which is then subtracted from the SE distribution to obtain the yield distribution of the trigger--correlated jet population.

The resulting distribution is still smeared in \pTjetReco\ due to detector effects and residual background fluctuations, which are corrected by iterative Bayesian unfolding~\cite{DAgostini:1994fjx}. Residual background fluctuations are evaluated using a data--driven method~\cite{STAR:2017hhs,STAR:2023pal,STAR:2023ksv}, while detector effects are simulated based on GEANT3~\cite{Brun:1987ma} and a detailed model of the STAR apparatus~\cite{STAR:2002eio}. The most significant detector effects are TPC tracking efficiency and resolution. In constructing the response matrix, the selection of $\pTjet>4$ \gev\ is applied to both reconstructed and simulated jets to avoid possible model inaccuracies at low $p_{\rm T}$. The prior distribution is generated by PYTHIA 6.428 (STAR tune)~\cite{Sjostrand:2006za,STAR:2019yqm}, and the number of unfolding iterations is set to 4 as the variations in unfolded spectra between successive iterations become smaller than the statistical errors.

Evaluation of systematic uncertainties for both the di--hadron and recoil jet \Icp\ measurements includes variation of event $v_{z}$ range, EA boundaries, and magnetic field direction. For di--hadrons, additional systematic variation includes track DCA selection, yield integration range, background fitting range, and $v_{n}$ values. For recoil jets, additional systematic variation includes lower and upper \pT\ track limits, ME normalization, tracking efficiency, choice of prior, and number of unfolding iterations. Correlations of uncertainties in the high--EA and reference EA intervals are accounted for, and the total \Icp\ uncertainty is the quadrature sum of the individual components. For di--hadrons the systematic uncertainty is less than 4\% for near--side and less than 5\% for recoil, while for recoil jets it varies between 2.3 and 7.3\%.

\begin{figure}[btph]
\begin{center}
\includegraphics[width=0.9\columnwidth, trim=25 0 20 30, clip]{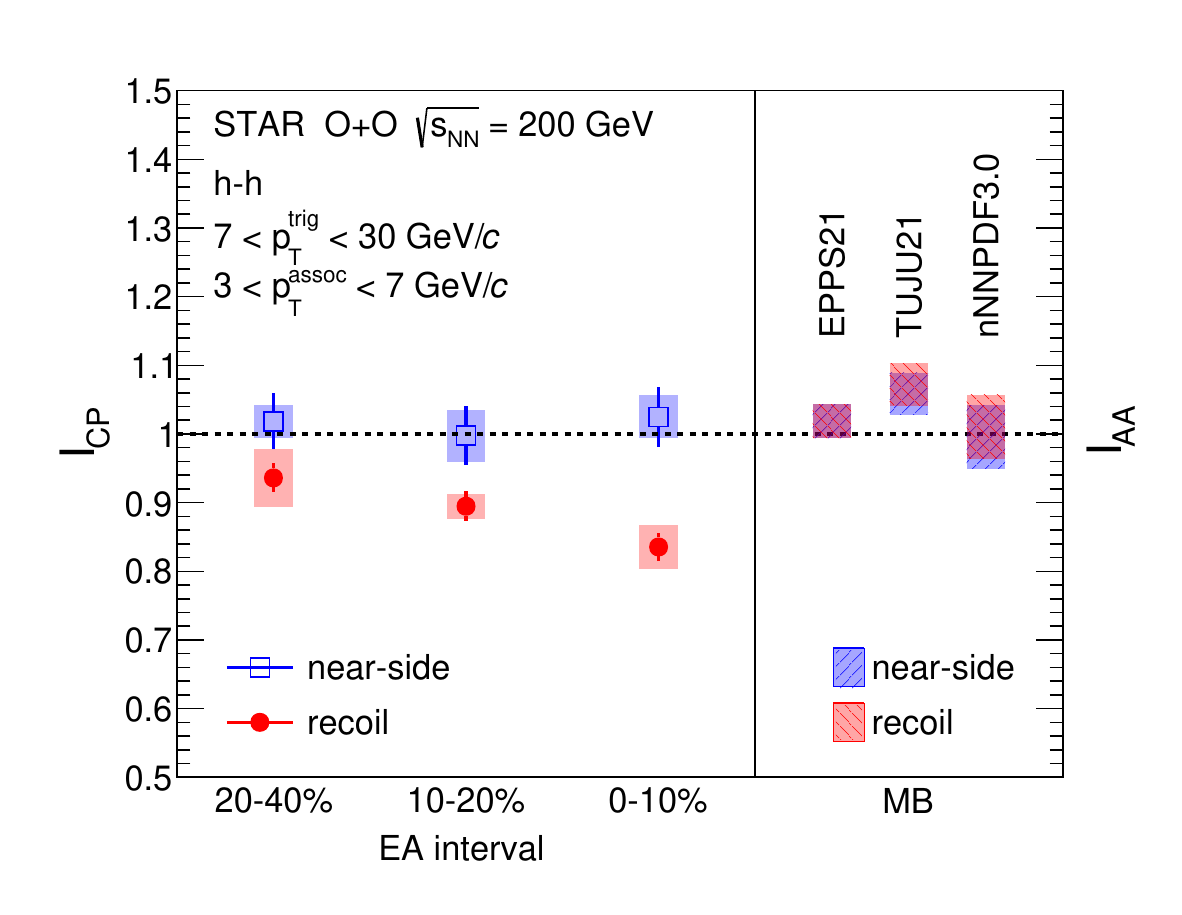}
\includegraphics[width=0.9\columnwidth, trim=25 25 20 30, clip]{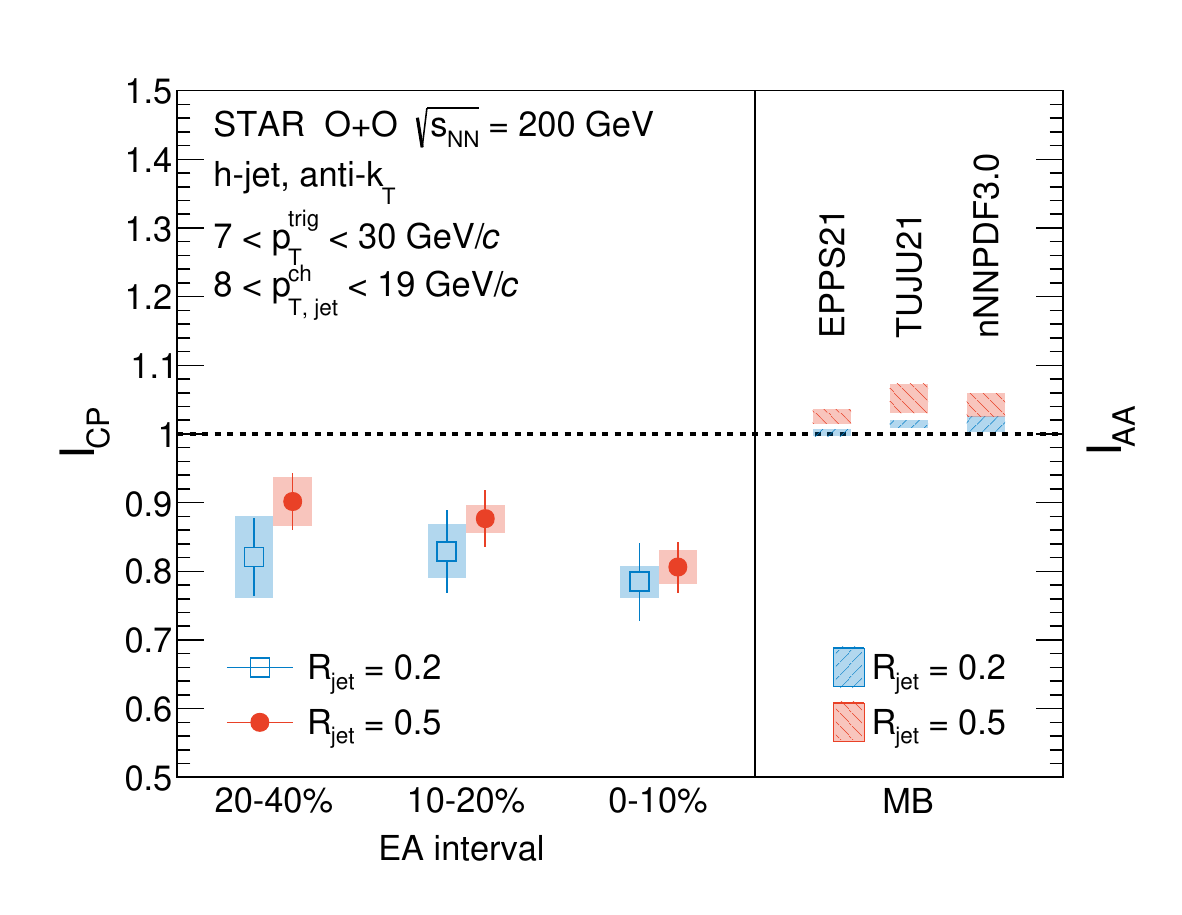}
\caption{Left panels: \Icp\ for associated hadrons in near--side and recoil peaks (upper) and recoil jets (lower) as a function of EA, with 40-60\% EA population used as reference. Vertical bars and shaded boxes are statistical errors and systematic uncertainties, respectively. Right panels: corresponding ``no quenching'' pQCD calculations for MB O+O collisions incorporating nPDF effects~\cite{Gebhard:2024flv}. Bands indicate 68\% confidence interval.}
\label{Fig:integrated_Icp}
\end{center}
\end{figure}

Figure \ref{Fig:integrated_Icp}, upper left panel, shows values of di--hadron \Icp\ for near--side and recoil associated charged hadrons, in three high--EA intervals. The near--side \Icp\ is consistent with unity, indicating that the jets generating trigger particles experience negligible energy loss~\cite{STAR:2006vcp}. In contrast, the recoil yield is suppressed, i.e. $\Icp<1$. For the 0-10\% EA interval, \Icp\ = $0.836\pm0.021\ (\rm stat.)\pm0.032\ (\rm syst.)$, significantly larger than that observed for similar measurements in Au+Au collisions~\cite{STAR:2006vcp}. The EA dependence of the recoil--side \Icp\ is fit with a linear function by taking 5\%, 15\% and 30\% as EA percentage values. The resulting slope is $-0.020\pm 0.012$.

Figure \ref{Fig:integrated_Icp}, lower left panel, shows recoil jet \Icp\ integrated over $8<p_{\rm T,jet}^{\rm ch}<19$ \gev, in three high--EA intervals. The recoil jet \Icp\ is consistent between $R=0.2$ and 0.5 within uncertainties. In the 0-10\% EA interval, recoil jet \Icp\ = $0.806\pm0.037\ (\rm stat.)\pm0.025\ (\rm syst.)$ for $R=0.5$. The recoil--jet \Icp\ as a function of EA is likewise fit with a linear function, and the resulting slopes are $-0.019 \pm 0.014$ for $R=0.5$ and $-0.008 \pm 0.020$ for $R=0.2$.

The suppression of recoil hadron or jet yield can arise from both initial--state shadowing effects and jet quenching. To estimate the shadowing effects, the right panels of Fig.~\ref{Fig:integrated_Icp} show theoretical calculations based on perturbative QCD (pQCD) for both recoil hadron and recoil jet, incorporating nuclear modification to PDFs (nPDFs) but not jet quenching~\cite{Gebhard:2024flv}. Three different nPDF parametrizations are employed~\cite{Gebhard:2024flv}, which do not incorporate spatial variation in nuclear modification due to lack of constraints from data~\cite{STAR:2017enh,STAR:2021wwq,ALICE:2021tyx}. The calculations therefore utilize the MB O+O yield normalized by the corresponding yield in \pp\ collisions, denoted \IAA. These calculations are thus not precisely comparable to measurements. Nevertheless, they provide the most direct assessment currently available of nPDF effects on the measured observables, and are therefore utilized here for guidance.

For di--hadron correlations, nPDF effects result in a value of \IAA\ a few percent greater than unity, for both near--side and recoil correlations. The three nPDF parameterizations are largely consistent. The calculations are likewise consistent with the measurement of near--side \Icp\ but are incompatible with the measured recoil \Icp, which is significantly suppressed below unity. The observation that the nPDF calculations predict similar effects for near--side and recoil hadron correlations suggests that the near--side measurement provides a suitable, data--driven no--quenching baseline. For semi--inclusive h+jet correlations, the calculations incorporating nPDFs are similarly incompatible with the observed strong suppression of the recoil jet yield.

\begin{table}[h!]
\caption{Significance of recoil yield suppression in h-h and h-jet correlations in the 0-10\% EA population, compared to several different choices of no--quenching baseline. See text for details.}
\centering
 \begin{tabular}{ |c||c |c |c|} 
 \hline
 Baseline & h-h recoil & h-jet, $R=0.5$ & h-jet, $R=0.2$ \\ 
  \hline
Unity & 4.3$\sigma$ & 4.3$\sigma$ & 3.5$\sigma$ \\
h-h near side  & 3.3$\sigma$ & N/A & N/A \\
EPPS21  & 4.0$\sigma$ & 4.8$\sigma$ & 3.5$\sigma$ \\
TUJU21  & 5.2$\sigma$ & 4.9$\sigma$ & 3.7$\sigma$ \\
nNNPDF3.0  & 2.7$\sigma$ & 5.0$\sigma$ & 3.7$\sigma$ \\
\hline
\end{tabular}
\label{Tab:sig}
\end{table}

To quantify the significance of the observed recoil yield suppression, Table~\ref{Tab:sig} presents the difference between the measurements and several different choices of ``no quenching'' baseline, expressed as a multiple of the uncertainty $\sigma$. The baseline choices are unity, the near--side di--hadron correlation, and the pQCD calculation with each of the nPDF parameterizations. The value of $\sigma$ is the quadrature sum of total uncertainties in data and the baseline if any. When using the near--side h-h \Icp\ as the baseline, its correlation in systematic uncertainties to those of the recoil--side \Icp\ has been accounted for. Table~\ref{Tab:sig} shows that the yield suppression has significance between $2.7\sigma$ and $5.2\sigma$, depending on the observable and choice of baseline. The suppression is significant, even when considering nPDF effects. We therefore conclude that the reported measurements provide strong evidence of jet quenching in high--EA O+O collisions at $\sqrtsnn=200$ GeV. 

Finally, the recoil jet yield suppression in Fig.~\ref{Fig:integrated_Icp} can be expressed in terms of a spectrum shift using the phenomenological analysis outlined in Ref.~\cite{ALICE:2017svf} to account for the shape of the recoil jet spectrum. Slopes of recoil jet spectra, based on exponential fits and averaged across all EA classes, are $3.16\pm0.12~(\rm stat.)$ for $R=0.2$ and $3.25\pm0.08~(\rm stat.)$ for $R=0.5$. The resulting effective jet \pT\ shift for the 0-10\% EA collisions is $0.77\pm0.23~(\rm stat.)~\pm0.09~(\rm syst.)$  GeV/$c$ for $R=0.2$ and $0.70\pm0.15~(\rm stat.)~\pm0.10~(\rm syst.)$ GeV/$c$ for $R=0.5$, which are consistent within uncertainties. This quantifies jet energy redistribution due to final--state interactions, and provides insight into its radial dependence relative to the jet axis. The $R$--dependence of the shift is consistent with that observed in Au+Au collisions~\cite{STAR:2017hhs}  within uncertainties, though the central value is smaller in O+O collisions.

In summary, the STAR experiment at RHIC presents new measurements that probe jet quenching in O+O collisions at \sqrtsnn\ = 200 GeV, based on di--hadron and semi--inclusive hadron--jet correlations. Jet quenching is quantified by comparing yields in high and low EA populations. Significant suppression of recoil di--hadron yield is observed in high--EA events, with no modification observed in the near--side di--hadron yield. The semi--inclusive recoil jet yield is likewise observed to be suppressed, with similar suppression magnitude observed for $R=0.2$ and 0.5. 
Model calculations incorporating nuclear PDF effects are insufficient to explain the recoil yield suppression. These measurements quantify jet quenching effects in O+O collisions, the smallest collision system in which this phenomenon has been observed. They provide new insight into the system--size dependence of jet quenching, and the mechanisms underlying the interaction of energetic partons with the quark--gluon plasma. 

\textit{Acknowledgments} We thank Jannis Gebhard, Aleksas Mazeliauskas, and Adam Takacs for providing theoretical calculations. We thank the RHIC Operations Group and SCDF at BNL, the NERSC Center at LBNL, and the Open Science Grid consortium for providing resources and support.  This work was supported in part by the Office of Nuclear Physics within the U.S. DOE Office of Science, the U.S. National Science Foundation, National Natural Science Foundation of China, Chinese Academy of Science, the Ministry of Science and Technology of China and the Chinese Ministry of Education, NSTC Taipei, the National Research Foundation of Korea, Czech Science Foundation and Ministry of Education, Youth and Sports of the Czech Republic, Hungarian National Research, Development and Innovation Office, New National Excellency Programme of the Hungarian Ministry of Human Capacities, Department of Atomic Energy and Department of Science and Technology of the Government of India, the National Science Centre and WUT ID-UB of Poland, German Bundesministerium f\"ur Bildung, Wissenschaft, Forschung and Technologie (BMBF), Helmholtz Association, Ministry of Education, Culture, Sports, Science, and Technology (MEXT), Japan Society for the Promotion of Science (JSPS), and Agencia Nacional de Investigacion y Desarrollo de Chile (ANID), Chile.

\textit{Data availability} The data that support the findings of this article are openly available~\cite{hepdata.182259}.

\bibliography{references}

\section*{End Matter}
\label{sect:SM}

\subsection*{Ratio of \epdmip\ distributions}
\begin{figure}[btph]
\begin{center}
\includegraphics[width=0.9\columnwidth, trim=10 15 40 20, clip]{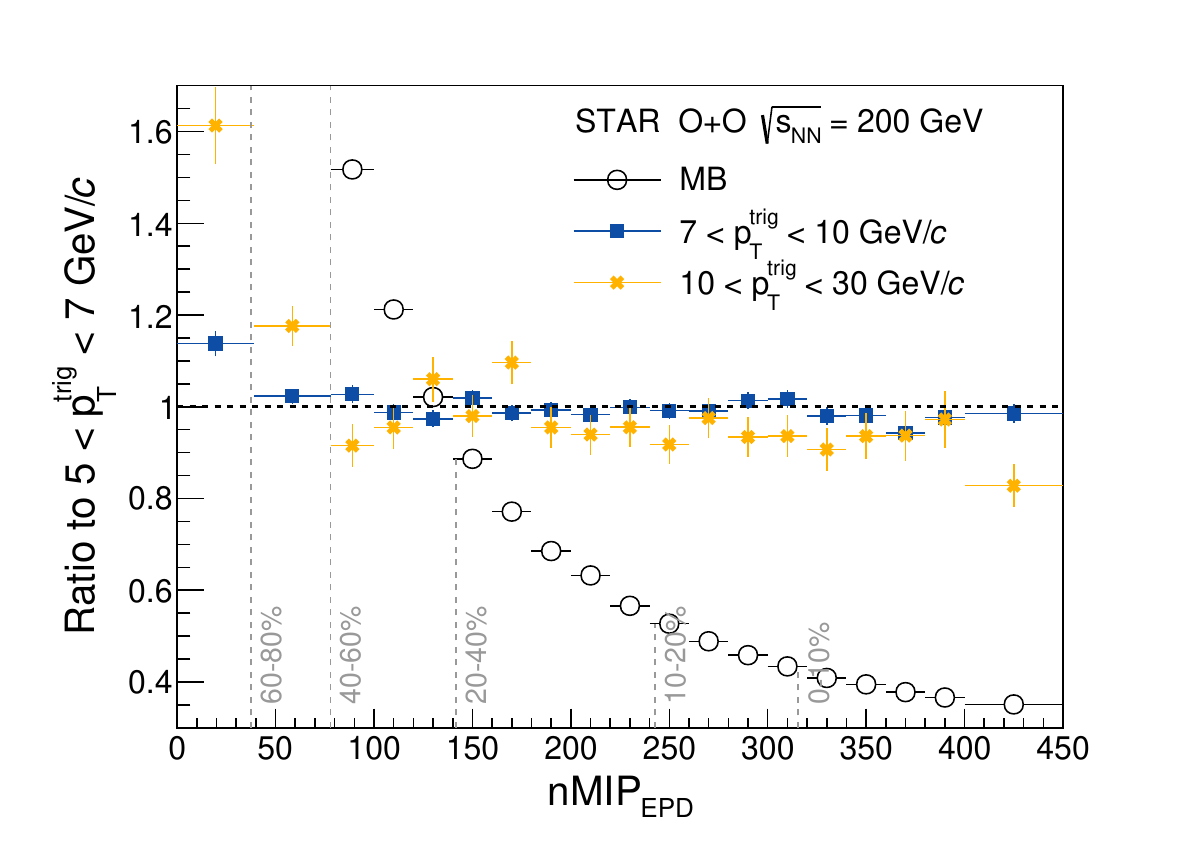}
\caption{Ratios of self-normalized \epdmip\ distributions for MB events, events with hadron triggers within 7 $<$ \pTtrig\ $<$ 10 \gev\ and  10 $<$ \pTtrig\ $<$ 30 \gev\ to that of 5 $<$ \pTtrig\ $<$ 7 \gev. Vertical bars around data points, which are mostly smaller than the marker size, indicate statistical errors.}
\label{Fig:EA ratio}
\end{center}
\end{figure}

Ratios of self-normalized \epdmip\ distributions for various event selections, including MB events, events with hadron triggers within 7 $<$ \pTtrig\ $<$ 10 \gev\ and  10 $<$ \pTtrig\ $<$ 30 \gev, to that of events with 5 $<$ \pTtrig\ $<$ 7 \gev\ are shown in Fig. \ref{Fig:EA ratio}. While a clear difference in the \epdmip\ distribution between MB events and events with high--\pT\ triggers is seen, the \epdmip\ distribution is largely invariant against hadron trigger \pT\ for \epdmip\ $>$ 78.

\subsection*{$f^{\rm ME}$ determination}

Tables \ref{tab:EA_bins_R02} and \ref{tab:EA_bins_R05} list $f^{\rm ME}$ values and associated uncertainties for $R=0.2$ and 0.5 jets, respectively, in different EA intervals. \pTjetReco\ ranges used for determining them are also tabulated. 
\begin{table}[h!]
\caption{\label{tab:EA_bins_R02} \pTjetReco\ ranges of $R=0.2$ jets, in unit of \gev, for determining $f^{\rm ME}$ values and associated uncertainties in different EA intervals. Resulting $f^{\rm ME}$ with statistical errors (first) and systematic uncertainties (second) are also listed.}
\begin{tabular}{ c|c c c} 
\hline
EA interval 
& \begin{tabular}{c}
\pTjetReco\ range \\
(default)
\end{tabular}
& \begin{tabular}{c}
\pTjetReco\ range \\
(variation)
\end{tabular}
& $f^{\rm ME}$ \\
\hline
0-10\% & (-1, 0) & (-1, 1) & $0.92\pm0.01\pm0.13$ \\
10-20\% & (-1, 0) & (-1, 1) & $0.85\pm0.01\pm0.23$ \\
20-40\% & (-1, 0) & (-1, 1) & $0.85\pm0.02\pm0.32$ \\
40-60\% & (-1, 1) & (-1, 0), (-1, 2) & $1.33\pm0.01\pm0.06$ \\
60-80\% & (-1, 1) & (-1, 0), (-1, 2) & $1.64\pm0.01\pm3.04$ \\
\hline
\end{tabular}
\end{table}

\begin{table}[h!]
\caption{\label{tab:EA_bins_R05} \pTjetReco\ ranges of $R=0.5$ jets, in unit of \gev, for determining $f^{\rm ME}$ values and associated uncertainties in different EA intervals. Resulting $f^{\rm ME}$ with statistical errors (first) and systematic uncertainties (second) are also listed.}
\begin{tabular}{ c|c c c} 
\hline
EA interval 
& \begin{tabular}{c}
\pTjetReco\ range \\
(default)
\end{tabular}
& \begin{tabular}{c}
\pTjetReco\ range \\
(variation)
\end{tabular}
& $f^{\rm ME}$ \\
\hline
0-10\% & (-3, -1) & (-3, -2), (-3, 0) & $0.55\pm0.02\pm0.08$ \\
10-20\% & (-3, -1) & (-3, -2), (-3, 0) & $0.54\pm0.04\pm0.46$ \\
20-40\% & (-2, -1) & (-2, 0) & $0.54\pm0.01\pm0.13$ \\
40-60\% & (-2, 0) & (-2, -1), (-2, 1) & $0.53\pm0.01\pm0.37$ \\
60-80\% & (-1, 0) & (-1, 1)& $0.42\pm0.01\pm0.38$ \\
\hline
\end{tabular}
\end{table}

\subsection*{Systematic uncertainties}

Systematic uncertainties for different sources are estimated as the variations in the results (yields and \Icp) by varying relevant analysis procedures or cuts. Specifically, the following changes are made:
\begin{itemize}
    \item{Common to h-h and h-jet measurements}
    \begin{itemize}
        \item $v_{z}$ range: $|v_{z}|<30$ cm (default), $|v_{z}|<15$ cm, $0 < v_z < 30$ cm, $-30 < v_z < 0$ cm
        \item EA boundaries: changing \epdmip\ boundaries such that an EA interval varies by 5\% relatively
        \item Magnetic field direction: analyzing data recorded with the magnetic filed aligning along $+z$ and $-z$ directions separately, while the combined results are taken as the default
    \end{itemize}

    \item{Specific to h-h measurements}
    \begin{itemize}
        \item DCA cut: 1.0 cm (default), 0.8 cm and 1.5 cm
        \item $v_{n}$ variation: $v_{2}$ = 0.05 (default), 0, 0.1; $v_{3}$ = 0.05 (default), 0, 0.1
        \item Background fit range: $0.8<|\Delta\phi|<1.6$ (default), $0.9<|\Delta\phi|<1.5$, $-1.6<\Delta\phi<-0.8$, $0.8<\Delta\phi<1.6$
        \item Signal extraction range: for near--side, $|\Delta\phi|<0.8$ (default), $|\Delta\phi|<\pi/2$, $|\Delta\phi|<0.7$; for recoil--side, $|\Delta\phi-\pi|<1$ (default), $|\Delta\phi-\pi|<\pi/2$, $|\Delta\phi-\pi|<0.9$
    \end{itemize}

    \item{Specific to h-jet measurements}
    \begin{itemize}
        \item Tracking efficiency: for primary tracks, varying the tracking efficiency by 5\% relatively; for secondary tracks, using the difference between 0\% efficiency and that of primary tracks
        \item Track \pT\ cut for event mixing: $p_{\rm T} > 5$ \gev\ (default), $p_{\rm T} > 4$ \gev, $p_{\rm T} > 7$ \gev
        \item $f^{\rm ME}$: see Tables \ref{tab:EA_bins_R02} and \ref{tab:EA_bins_R05}. Its total uncertainties are calculated by adding statistical errors and systematic uncertainties in quadrature. 
        \item Response matrix \pTjet\ range: \pTjet\ $> 4$ \gev\ (default), \pTjet\ $> 3$ \gev, \pTjet\ $> 5$ \gev
        \item Unfolding: flatter or steeper prior; number of iterations changed from default 4 to 3 and 5
    \end{itemize}
    
\end{itemize}

If multiple variations are made for one source, the root mean square of the changes in the results is taken as the corresponding uncertainty. The total uncertainties are obtained as the quadrature sum of the individual contributions. Tables \ref{tab:sys_hh} and \ref{tab:sys_hjet} list individual and total uncertainties for the h-h and h-jet \Icp\ in the 0-10\% EA interval, respectively. 

\begin{table}[h!]
\caption{\label{tab:sys_hh} Relative individual and total systematic uncertainties for near--side and recoil--side h-h \Icp\ in 0-10\% EA interval.}
\begin{tabular}{ c|c c} 
\hline
Sources  & near--side & recoil \\
\hline
DCA cut & 0.5\% & 0.1\% \\
$v_{n}$ variation & 0.4\% & 0.2\% \\
Background fit range & 2.8\% & 1.3\% \\
Signal extraction range & 0.4\% & 0.2\%\\
$v_{z}$ range & $<0.1\%$ & 3.0\% \\
EA boundaries & 1.0\% & 2.1\% \\
Magnetic field direction & $<0.1\%$ & $<0.1\%$  \\
\hline
\hline
Total & 3.1\% & 3.9\%\\
\hline
\end{tabular}
\end{table}

\begin{table}[h!]
\caption{\label{tab:sys_hjet} Relative individual and total systematic uncertainties for $R=0.2$ and 0.5 h-jet \Icp\ in 0-10\% EA interval.}
\begin{tabular}{ c|c c} 
\hline
Sources  & $R=0.2$ & $R=0.5$ \\
\hline
Tracking efficiency & 0.8\% & 0.1\% \\
Track \pT\ cut for event mixing & 0.1\% & 0.1\% \\
$f^{\rm ME}$ & 0.1\% & 0.3\%\\
Response matrix \pTjet\ range & 0.2\% & 0.2\% \\
Unfolding & 0.3\% & 0.7\% \\
$v_{z}$ range & 0.4\% & 0.6\% \\
EA boundaries & 1.5\% & 1.0\% \\
Magnetic field direction & $2.3\%$ & 2.8\% \\
\hline
\hline
Total & 2.9\% & 3.1\%\\
\hline
\end{tabular}
\end{table}

\subsection*{h-jet yields and \Icp}
\begin{figure}[btph]
\begin{center}
\includegraphics[width=0.85\columnwidth, trim=5 45 5 45, clip]{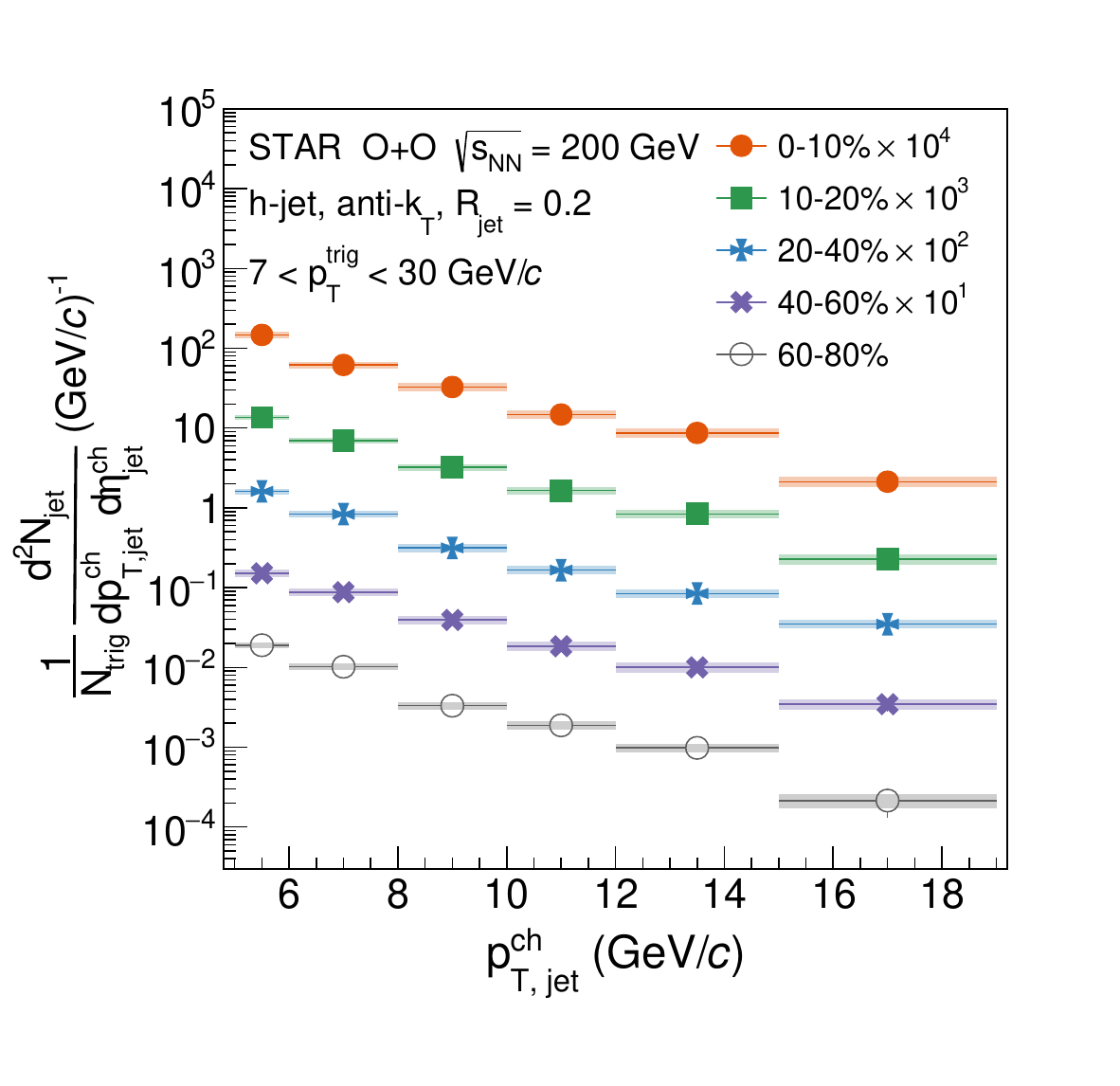}
\includegraphics[width=0.85\columnwidth, trim=5 45 5 45, clip]{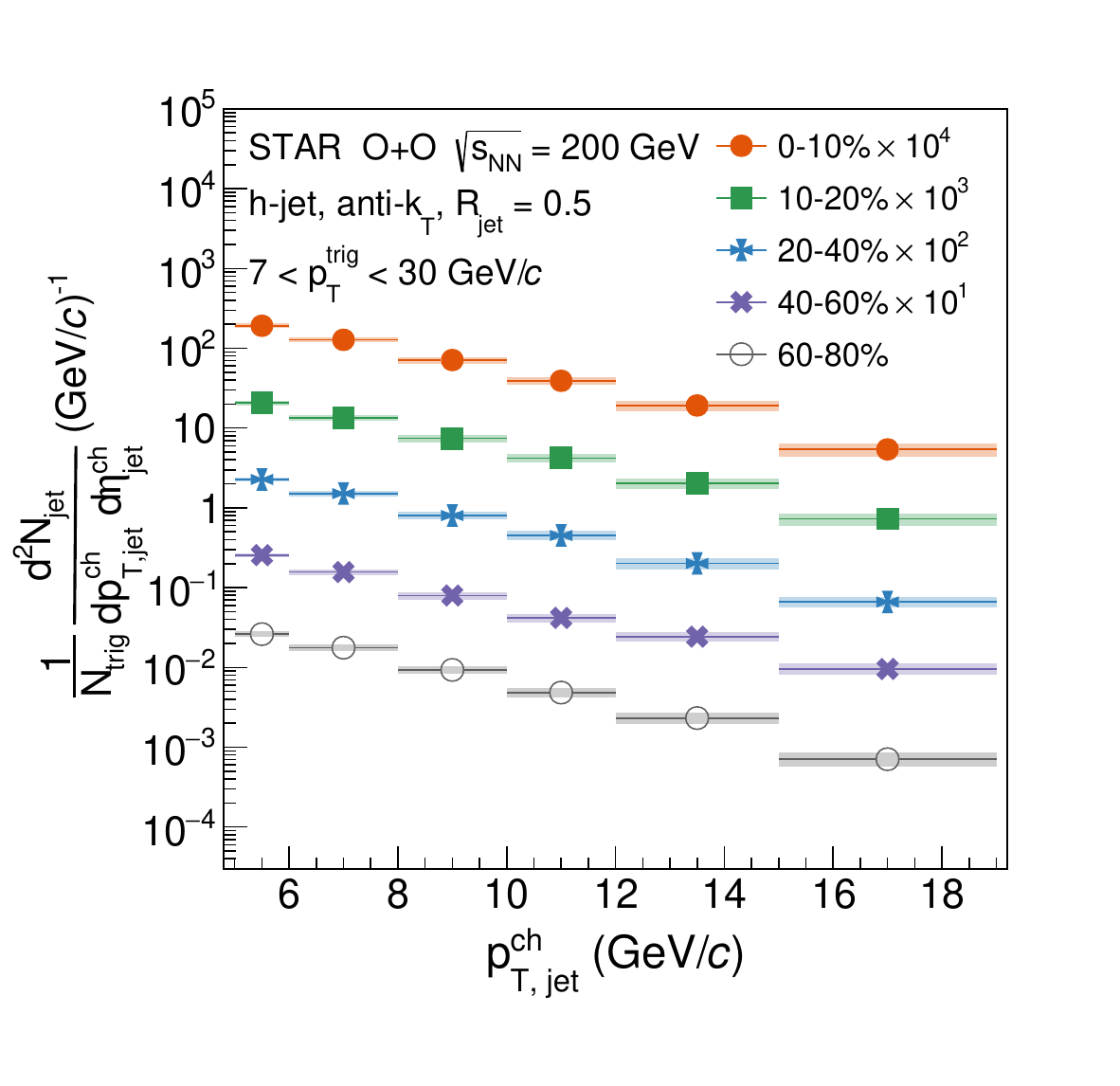}
\caption{Recoil jet spectra as a function of \pTjet\ for jet radii of $R=0.2$ (top) and 0.5 (bottom) in different EA classes of O+O collisions at \sqrtsnn\ = 200 GeV. Vertical bars and shaded boxes around data points display statistical errors and systematic uncertainties, respectively.}
\label{Fig:hjet_yield}
\end{center}
\end{figure}

Fully-corrected recoil jet yields per trigger as a function of \pTjet\ in O+O collisions at \sqrtsnn\ = 200 GeV are shown in Fig. \ref{Fig:hjet_yield} for jet radii $R=0.2$ and 0.5. Distributions in different EA classes are shifted vertically for clarity. 

\begin{figure}[btph]
\begin{center}
\includegraphics[width=0.85\columnwidth, trim=25 20 25 25, clip]{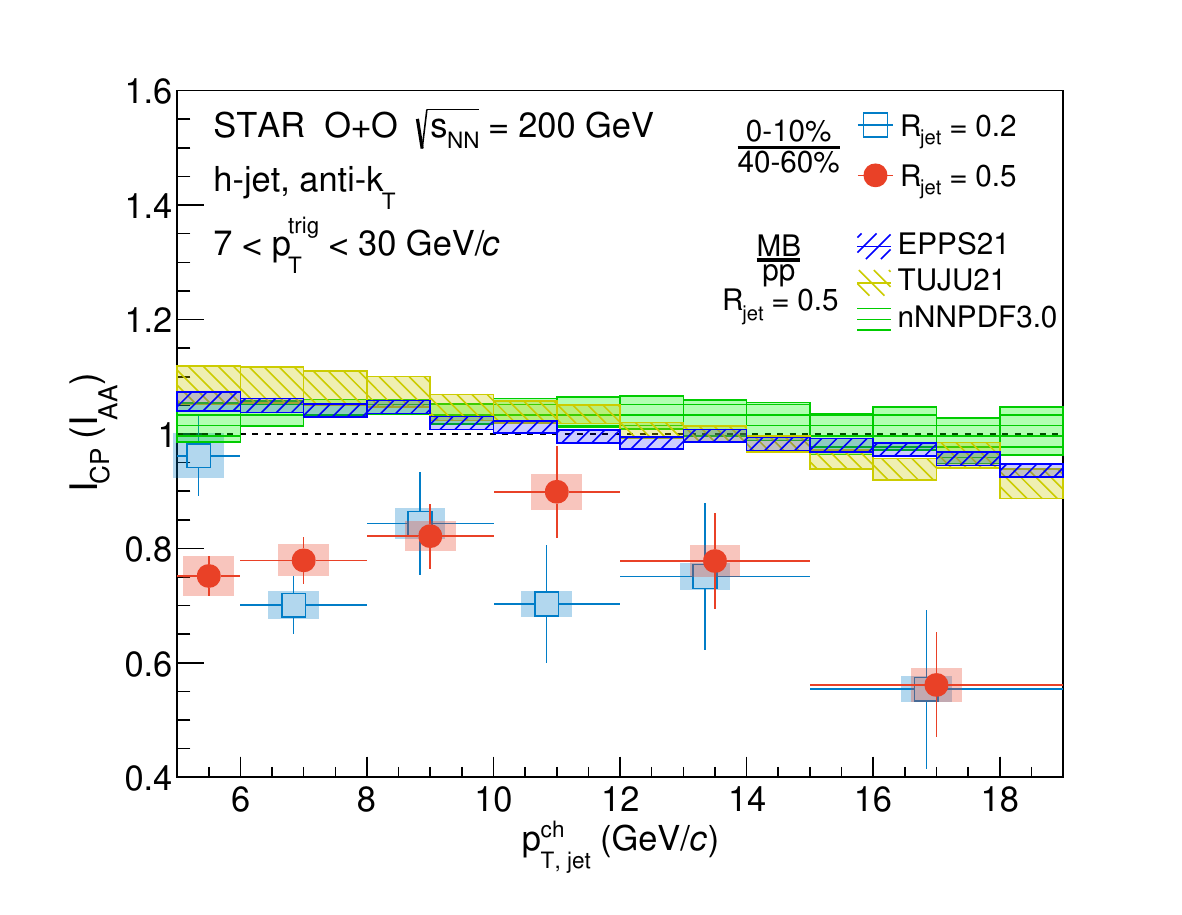}
\caption{Recoil jet \Icp\ between 0--10\% and 40--60\% EA classes as a function of \pTjet\ for jet radii of $R=0.2$ and 0.5. Vertical bars and shaded boxes around data points display statistical errors and systematic uncertainties, respectively. Data points for $R=0.2$ are shifted to the left slightly for clarity.}
\label{Fig:hjet}
\end{center}
\end{figure}

Figure \ref{Fig:hjet} shows \Icp, the ratio of fully corrected per--trigger recoil jet yields for the 0--10\% and 40--60\% EA classes, as a function of \pTjet. A marked suppression below unity is observed, with no significant dependence on \pTjet.

\end{document}